\documentclass[a4paper,12pt]{paper}
\usepackage[latin1]{inputenc}
\usepackage{exscale}
\usepackage{graphicx}
\usepackage{amsfonts}
\usepackage{amsmath}
\usepackage[left=35mm,top=25mm,right=30mm,textheight=230mm,textwidth=135mm]{geometry}
\newcommand {\GeV}{\,\mathrm{GeV}}
\newcommand{\I}{{\it I}}
\title{A Multi-variate Discrimination Technique Based on Range-Searching}

\author{T. Carli$^{\,a}$ and B. Koblitz$^{\,b}$}

\institution{$^a$ DESY, Hamburg and Universit\"at Hamburg\\
$^b$ MPI f\"ur Physik, M\"unchen}

\begin{document}

\maketitle
\begin{abstract}
  We present a fast and transparent multi-variate event classification
  technique, called PDE-RS, which is based on sampling the signal and
  background densities in a multi-dimensional phase space using
  range-searching.  The employed algorithm is presented in detail and
  its behaviour is studied with simple toy examples representing
  basic patterns of problems often encountered in High Energy Physics
  data analyses. In addition an example relevant for the search for
  instanton-induced processes in deep-inelastic scattering at HERA is
  discussed.  For all studied examples, the new presented method
  performs as good as artificial Neural Networks and has furthermore
  the advantage to need less computation time. This allows to
  carefully select the best combination of observables which optimally
  separate the signal and background and for which the simulations
  describe the data best. Moreover, the systematic and statistical
  uncertainties can be easily evaluated. The method is therefore a
  powerful tool to find a small number of signal events in the large
  data samples expected at future particle colliders.
\end{abstract}


\begin{keywords}
  probability density estimation, multi-variate discrimination
  technique, range-searching, event classification, Neural Networks,
  instanton-induced processes, deep-inelastic scattering, HERA
\end{keywords}

\section{Introduction}
In High Energy Physics one is frequently confronted with the task of
finding a small number of distinctive (signal) events among a large
number of background events. This problem is often tackled by
simply applying cuts, motivated by models describing these characteristic
events, on measured characteristic observables. However, in particular
in complex cases where many observables have to be used, more powerful
techniques exist, which are based on probability density estimation
(PDE) or which employ Neural Networks (NNs). They first combine the
observables to a single one, called ``discriminant'' on which then a
cut to separate signal from background is applied. These methods were
widely used in the search for the top-quark at the TEVATRON
\cite{Bhat1,Holmstroem,Raja} and
will play a major role in the search for the Higgs boson in the 
future \cite{Bhat2}. For a general introduction to multi-variate
discrimination techniques see \cite{Bishop}.
 
A drawback of these methods is often that they come as a ``black box''
which provides few insights on the statistical and systematic
uncertainties of the results obtained with the method. We will present here a novel
discrimination technique based on counting signal and background
events in small multi-dimensional boxes. The simple event counting
allows to transparently handle the involved uncertainties and to
separate signal and background events with a power similar to NNs. The
event counting is done using a fast range-searching algorithm. Its
speed allows to use very large data samples required in
analyses where a high reduction of background is necessary or to scan
a large number of observables for those which give the best separation
of signal and background and for which the Monte Carlo simulations
describe the data best. This technique has already been used in recent
searches at HERA \cite{H1Instantons,ZeusTau}.

\section{Probability Density Estimation Techniques}
In order to classify an event it is necessary to estimate the
probability $p(\mathbf x)$ that an event is of the signal class, given
$d$ measured properties (called observables in the following) $\mathbf
x=(x_1,\ldots,x_d)$. An estimate $\tilde p(\mathbf x)$ of $p(\mathbf
x)$ can be obtained by employing Monte Carlo simulators which
approximate the probability densities of the signal $\tilde
\rho_s(\mathbf x)\approx\rho_s(\mathbf x)$ and of the background
events $\tilde \rho_b(\mathbf x)\approx\rho_b(\mathbf x)$ by sampling
the $d$-dimensional phase space with simulated events. The probability
that a particular event belongs to the signal class is then given by
\begin{equation}
  \label{eq:probdef}
  \tilde p(\mathbf x)=\frac{\tilde \rho_s(\mathbf x)}{\tilde
    \rho_s(\mathbf x)+\tilde \rho_b(\mathbf x)}\approx
  p(\mathbf x)=\frac{\rho_s(\mathbf x)}{\rho_s(\mathbf x)
    +\rho_b(\mathbf x)}.
\end{equation}
The function $\tilde p(\mathbf x)$ is a so called ``Discriminant'', since
it assigns to any given combination of measured observables a single
value which discriminates background from signal events. 

Finding good approximations of the signal and background densities can
be a rather difficult problem, especially in high dimensional cases.
Here, histograming methods cannot be used because the number of
required bins increases as $m^d$, if $m$ is the number of bins per
dimension. This causes a dramatic increase in memory usage and a
decrease of the available number of events per bin.  The problem is
aggravated by correlations among the observables which is often the
case in High Energy Physics applications. Due to correlations, the
phase space is commonly populated only in a sub-space of lower
dimensionality, i.e. the {\it intrinsic dimensionality} of the problem
actually is smaller. To overcome the problem of the high
dimensionality, sometimes methods are employed, which try to deduce
the multi-dimensional probability density from projections. These
methods suffer from correlations among the observables, which are not
modelled by the projections.

Kernel based PDE methods \cite{Holmstroem} sum up appropriately chosen
{\it kernel} functions to model the probability density around the
point $\mathbf x$:
\begin{equation}
  \label{eq:kernelEstimation}
  \tilde p(\mathbf x)=\frac{1}{Nh^d} \sum_{i=1}^N K\left(\frac{\mathbf
    x-\mathbf x_i}{h}\right),
\end{equation}
where the sum runs over all sample events $\mathbf x_i$, $N$ is the total
number of events in the data sample and $h$ is a
smoothing parameter. For the kernel function $K$ often a
Gaussian distribution is chosen:
\begin{equation}
  K(\mathbf x)=\frac{1}{(2\pi)^{d/2}}\exp\left(-\frac{1}{2}||\mathbf
    x||^2\right),
\end{equation}
making the resulting distribution continuous and differentiable. Since
for every event which is classified, eq.~(\ref{eq:kernelEstimation}) needs
to be evaluated involving the sum over all $N$ sample events, these
methods are very time consuming for large data samples.
To avoid this problem functions only defined in a small region
around $x$ have to be used.

A powerful method to estimate $p(\mathbf x)$ are artificial Neural
Networks \cite{Bishop}. Their design is inspired by biological
neurons. In a training phase they parameterise the probability density
by linear combinations of smooth functions. Using training events of
known type the free parameters, usually called weights, are adjusted.
The convergence of this procedure is usually fast reducing the time
requirements compared to kernel based PDE methods. Given sufficiently
large NNs with a high number of nodes, even very complicated
probability densities can be approximated.  Their good performance and
their fast applicability make them a good candidate to compare the new
method with. In many problems a fast computing time is crucial to
perform a meaningful reduction of input observables by studying
combinations of them and is important to handle large data samples
required in searches where the background needs to be strongly
reduced.

\section{Probability Density Estimation based on Range-Searching}
\subsection{The PDE-RS Method}
The multi-variate probability density estimation technique based on
range-searching (PDE-RS) counts the number of Monte Carlo generated
signal and background events in the vicinity of an event which is to
be classified. From the counted events the probability of this event
to be of the signal class is derived.  This is done in an efficient
way using range-searching with an algorithm described below.  Given
the number of signal events $n_s$ and the number of background events
$n_b$ in a small volume $V(\mathbf x)$ around the point $\mathbf x$,
we define a discriminant
\begin{equation}
  \label{eq:defD}
  D(\mathbf x):=\frac{n_s}{n_s+c\, n_b},
\end{equation}
which for sufficiently small volumes $V(\mathbf x)$ and a sufficiently
high number of sample events gives a very good approximation of
$p(\mathbf x)$, if the normalisation constant $c$ is chosen such that
the total number of simulated signal events is equal to $c$ times the
total number of background events. $D(\mathbf x)$ provides a good
estimate of the local event density and prevents a wrong
classification due to bad interpolation into regions where the Monte
Carlo simulations provide no information. This happens, if methods try
to fit the event density globally, as is done e.g. by Neural Networks.
However, in the PDE-RS method a large number of Monte Carlo generated
events is needed to densely populate the whole phase space.  This can
be a limiting factor, if the number of observables and thus the
dimensionality of the problem is high.
For the counting method, in the vicinity of each event
to be classified a large number of events have to be counted --- a
potentially time-consuming task. This problem is known as
``range-searching'' in computer sciences.

Range-searching has been studied intensively since several years,
because the problem to find a specific event in a large data sample
occurs in all sorts of classification tasks.
Powerful algorithms have been devised to tackle it
\cite{Bentley,BentleyFriedman}. Two different classes of algorithms
are usually applied: One which subdivides the entire volume of the
observable space into small boxes and stores the events within the boxes
in a linked list\footnote{A data structure with a data element (the
  event) and a pointer to the next element.}. Searching for an event
then just involves looking up which boxes are in the vicinity of the
event that is to be classified and then simply scan the linked list
for events within a certain distance. The second class of algorithms
use multi-dimensional binary trees to store the events. An algorithm
of this class as described in \cite{Sedgewick} is used here. The
advantage is that in contrast to the subdivision algorithm the extent
of the observable space needs not to be known, that is the minima and
maxima of the observables need not be calculated before. In addition,
subdivision algorithms have a huge memory consumption if the
dimensionality of the problem is large, since they have to store the
pointers to the linked lists in an array of the dimension of the
problem, even if no event lies in a box.  Such a behaviour is, however, 
expected for High Energy Physics events, for which the observables describing
their properties are in many cases correlated leaving a large fraction
of the phase space empty.

\subsection{The PDE-RS Algorithm}
\begin{figure}
  \setlength{\unitlength}{1cm}
  \begin{picture}(13,5)
    \put(0,0){\includegraphics[width=\textwidth]{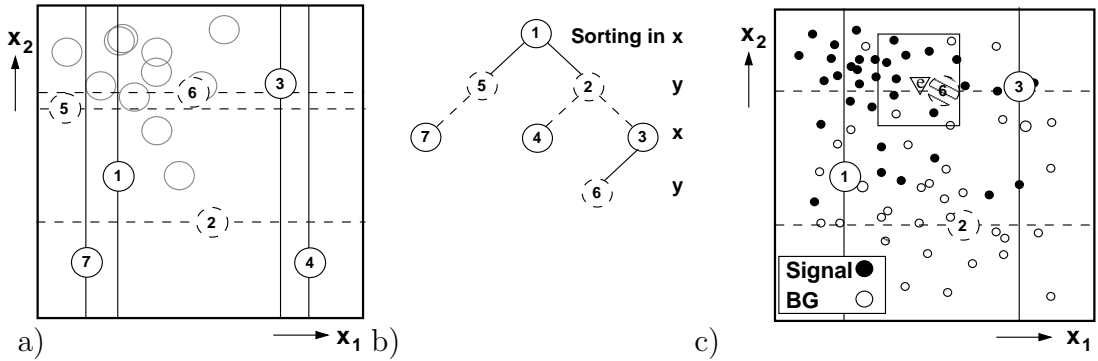}}
    \put(0,0){ a)}
    \put(4.7,0){ b)}
    \put(9,0){ c)}
  \end{picture}
  \caption{
    \label{fig:rsearch} a) 
    Signal events distributed in a $2$-dimensional phase space spanned
    by the observables $x_1$ and $x_2$. The numbers refer to the
    occurrence of the events in the data sample, b) the resulting
    binary tree to store the events and c) signal and background
    events in the $x_1$-$x_2$ plane. The open triangle depicts the
    event to be classified.  The box around the triangle shows the
    region where events are considered to calculate the discriminant.
    The lines and numbers illustrate the way how the events in the box
    are found with the help of the binary tree. A detailed description
    can be found in the text.  }
\end{figure}

The algorithm used for the event classification is based on the
range-searching algorithm described in
\cite{Sedgewick}\footnote{There, also some program code may be found.
  The full C++ implementation as it is used here is available from the
  authors ({\tt carli@mail.desy.de, koblitz@mail.desy.de}).}. It
allows to search through $N$ Monte Carlo generated events that sample
the signal and background density within a time $\sim \log_2(N)$. To
achieve this scaling of the algorithm with the total number of events,
all $N$ events are first stored in two $d$-dimensional binary trees
--- one for the background and one for the signal events --- as is
sketched in figure~\ref{fig:rsearch} for a two-dimensional example:
Consider a random sequence of signal events $e_i(x_1,x_2)$,
$i=1\ldots7$ shown in figure~\ref{fig:rsearch}a with their position in
$x_1-x_2$-space, which are to be stored in a binary tree. The first
event in the sequence becomes by definition the topmost node of the
tree. The second event $e_2(x_1,x_2)$ has a larger $x_1$-coordinate
than the first event, therefore a new node is created for it and the
node is attached to the first node as the right child (if the
$x_1$-coordinate had been smaller, the node would have become the left
child). Event $e_3$ has a larger $x_1$-coordinate than event $e_1$, it
therefore should be attached to the right branch below $e_1$. Since
$e_2$ is already placed at that position, now the $x_2$-coordinates of
$e_2$ and $e_3$ are compared, and, since $e_3$ has a larger $x_2$,
$e_3$ becomes the right child of the node with event $e_2$. Thus the
tree is sequentially filled by taking every event and, while
descending the tree, comparing its $x_1$ and $x_2$ coordinates with
the events already in place. Whether $x_1$ or $x_2$ are used to
compare depends on the level within the tree. On the first level,
$x_1$ is used, on the second level $x_2$, on the third again $x_1$ and
so on. The result for events $e_i$ is shown in
figure~\ref{fig:rsearch}b. The amount of time needed to fill the tree
is $\sim \sum_{i=1}^N \log_2(i)=\log_2(N!)=\mathcal{O}(N \log_2(N))$.  The
last equality can be easily verified with the help of {\sc Sterling}'s
formula.

Finding all events within the tree which lie in a given box is done
in a similar way by comparing the bounds of the box with the
coordinates of the events in the tree. For example, if the whole box
lies to the right of event $e_1$ as shown in
figure~\ref{fig:rsearch}c, then only events on the branch below and
including $e_2$ need to be searched. This halves the number of events
in question. Only if an event in a node lies within the bounds of the
coordinates of the box that it is compared to, both its siblings need
to be searched. Searching the tree once requires therefore an effort
only $\sim \log_2(N)$. It needs to be noticed that the whole tree of
Monte Carlo generated events needs to be kept in the main memory of
the computer to have a reasonably fast access time when comparing the
coordinates. Therefore, only the advent of computers with random
access memory of the order of hundreds of megabyte made it possible
to use millions of events to sample the signal and background
densities.

Since the number of signal and background events used in the
calculation of the probability density estimation is known
at each point in the phase space, the uncertainty of this
estimate due to the limited number of signal and
background events can be calculated. By inspecting
eq.~(\ref{eq:defD}) we find that the statistical error $\Delta D(\mathbf
x)$ is given by
\begin{equation}
  \label{eq:DUncertainty}
  \Delta D(\mathbf x)=\sqrt{\left(\frac{c\,n_b}{\left(n_s+c\,
          n_b\right)^2}
      \Delta n_s\right)^2 + \left(
  \frac{c\,n_s}{\left(n_s+c\, n_b\right)^2}\Delta
  n_b\right)^2},
\end{equation}
where $\Delta n_s$ and $\Delta n_b$ are the statistical uncertainties
of the signal and background respectively. 

Usually the calculated discriminant values of samples of events
are histogramed and the performance of the discrimination
technique is estimated by applying cuts on this discriminant. For
discriminant values falling into the same bin, the uncertainties are
of course correlated, because they are derived from the same samples of signal
and background events. Using the individual uncertainties on $D$ for
each event according to eq.~(\ref{eq:DUncertainty}) would overestimate
the systematic uncertainty\footnote{We call the uncertainty of the
  distribution of $D$ induced by the statistical uncertainties of the
  event samples used for the classification the systematic uncertainty
  of $D$.}. Instead, a good estimate of the systematic
uncertainty of each histogram bin is given by using the total number of
signal and background in eq.~(\ref{eq:DUncertainty}), which fall into the
phase space region corresponding to the bin. We have used Monte Carlo
experiments using statistically independent event samples to confirm
the validity of this estimate.

In order to influence the uncertainty of the probability density
estimate, the size of the box can be adapted. The lengths of the box
edges are the only free parameters of the algorithm. In the following
examples where the observables are produced in similar ranges (i.e.
they have similar ``scales''), the number of parameters is reduced by
using a hyper cube with edges of equal size $2l$. In this case $l$ is
the largest distance in the maximum norm of every counted event to the
centre of the box. In practical applications, where the observables
can have very different scales, the relative length of the edges can
be deduced from histograms of the observables before the method is
applied.  The box size should be chosen large enough in order to have
a reasonably small uncertainty on $D(\mathbf x)$. For too large boxes,
however, the precise mapping of the probability density onto boxes is
not possible and therefore the achievable separation is reduced.  As
will be shown in the following, the performance of the PDE-RS method
is not very sensitive with respect to the box size.

\section{Properties of PDE-RS and Comparison to Neural Networks}
In the following we will study the properties and the performance of
the PDE-RS method using three simple examples chosen as prototypes of
basic situations encountered in High Energy Physics data analyses.
The performance, the needed computing time and the ease of
applicability are compared to NNs.  The following examples have been
chosen:
\begin{enumerate}
\item Two Gaussian probability densities to study the simple case of
  uncorrelated observables, where most of the multi-variate
  discrimination techniques give good results.
\item Two strongly correlated observables, where an elementary
  variable transformation largely simplifies the classification
  problem, as example of a problem which can be easily solved, if the
  relations between the observables are known e.g. by insights into
  the acting physical laws. Such problems can easily be solved by a
  physicist when a small number of observables are involved, but can
  pose problems to classification algorithms.
\item A high dimensional example, where a physicist has difficulties
  to find a good separation by looking at the observable distributions
  and where the multi-variate discrimination techniques generally show
  their strength.
\end{enumerate}

\subsection{Bivariate Uncorrelated Gaussian Probability Densities}
\begin{figure}
  \setlength{\unitlength}{1cm}
  \begin{picture}(13,11.5)
    \put(0,0){\includegraphics[width=\textwidth]{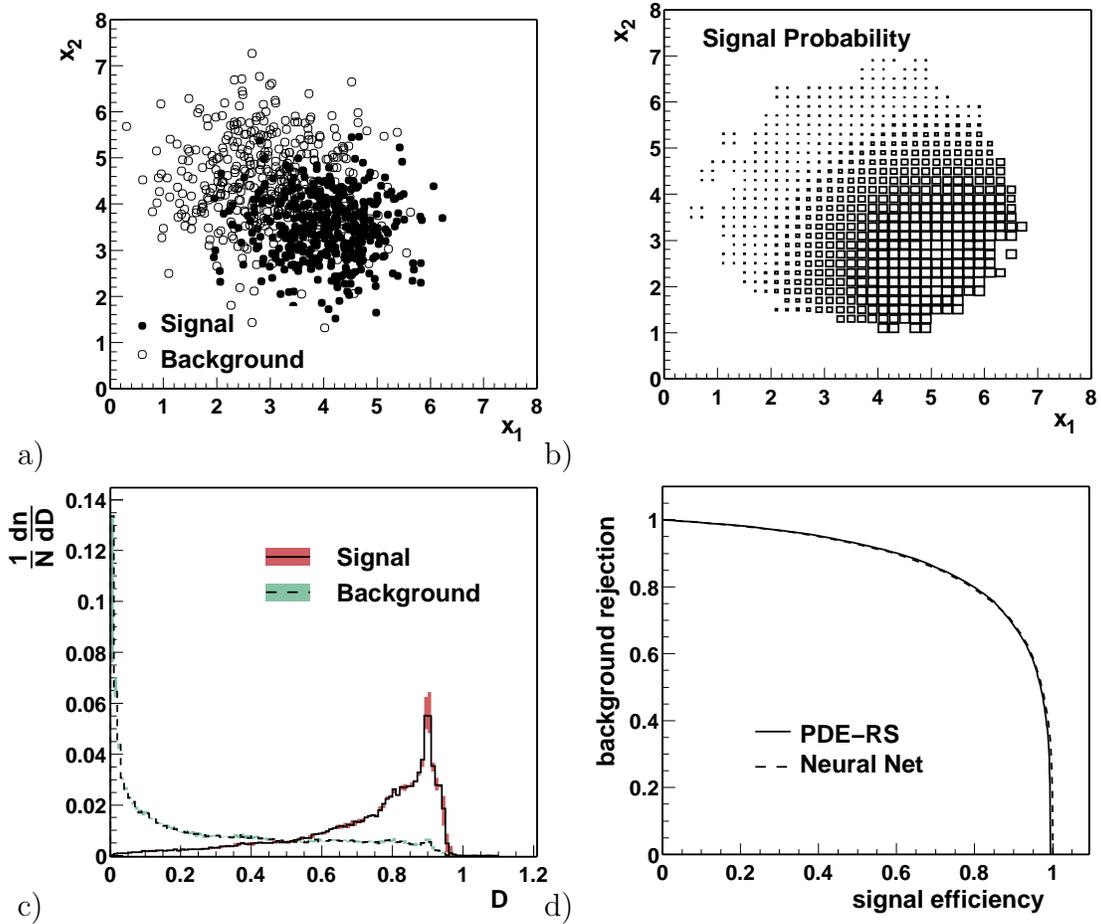}}
    \put(0,6.){ a)}
    \put(7,6.){ b)}
    \put(0,0){ c)}
    \put(7,0){ d)}
  \end{picture}
  \caption{
    \label{fig:twoGauss}
    a) The phase space densities of background (open circles) and
    signal (full circles) of two simple bivariate Gaussian
    distributions with displaced means and equal widths b) the
    probability density for signal type events estimated by the PDE-RS
    method c) the resulting shape of the discriminant distribution for
    signal and background events. The band indicates the statistical
    uncertainty of the discriminant.  d) background rejection, $ 1 -
    \epsilon_b$, versus signal efficiency for the PDE-RS method and a
    NN, obtained by cutting on the discriminant distributions.  }
\end{figure}
A very simple example using two two-dimensional Gaussians for the
background with means $\left<x_{1,b}\right>=3$ and
$\left<x_{2,b}\right>=4.5$ and widths of $\sigma_{1,b}=\sigma_{2,b}=1$
and for the signal with means $\left<x_{1,s}\right>=4$,
$\left<x_{2,b}\right>=3.5$ and widths of
$\sigma_{1,s}=\sigma_{2,s}=0.75$ is shown in
figure~\ref{fig:twoGauss}a. In figure~\ref{fig:twoGauss}b the
resulting probability density that an event is of signal type is
depicted. To calculate the probability density 100,000 events have
been filled into the two binary trees each and $\Delta D<0.05$ has
been required. A box of $V(x_1,x_2)=0.18\cdot0.18$ around each
classified event in which sample events are counted has been used.
The distributions of the discriminant $D$ of test signal and
background events\footnote{ In order not to bias the performance, the
  event samples are divided into one needed sample used to set-up the
  binary tree and one sample to test the performance of the
  classification algorithm.}  is shown in figure~\ref{fig:twoGauss}c.
Most of the background events are correctly classified and have a
small $D$ value.  Since there is no phase space region where there are
signal but no background events, the discriminant does not peak at $D
\approx 1$ but at a somewhat lower value. The shaded area depicts the
systematic uncertainty of the discriminant according to
eq.~(\ref{eq:DUncertainty}).  Finally, the background rejection
$1-\epsilon_b$, where $\epsilon_b$ is the background efficiency, is
shown as a function of the signal efficiency $\epsilon_s$ in
figure~\ref{fig:twoGauss}d and is compared to the result of a single
hidden layer feed forward NN\footnote{We used a modified version of
  the package written in C++ by J.~P.~Ernenwein, available at {\tt
    http://e.home.cern.ch/e/ernen/www/NN/}.} with 10 hidden nodes. In
an ideal case where every event is classified correctly, the area
below the background rejection -- signal efficiency curve would be a
square of area 1. For the PDE-RS method we find an area of $0.876\pm
0.01$, for the NN an area of 0.877. The two methods are compatible.
While the PDE-RS method performs slightly better for lower signal
efficiencies, its performance for $\epsilon_s\approx 1$ is slightly
below the one of the NN. The signal efficiency $\epsilon_s$ actually
never reaches 1, since there are always events which cannot be
classified due to the requirement that $D(\mathbf x)$ has to be
calculated with sufficient statistical precision, i.e. $\Delta
D<0.05$. The NN on the other hand classifies every event regardless of
the uncertainty of the fit to the probability density.

\subsection{Highly Correlated Observables}
\begin{figure}
  \setlength{\unitlength}{1cm}
  \begin{picture}(13,11.5)
    \put(0,0){\includegraphics[width=\textwidth]{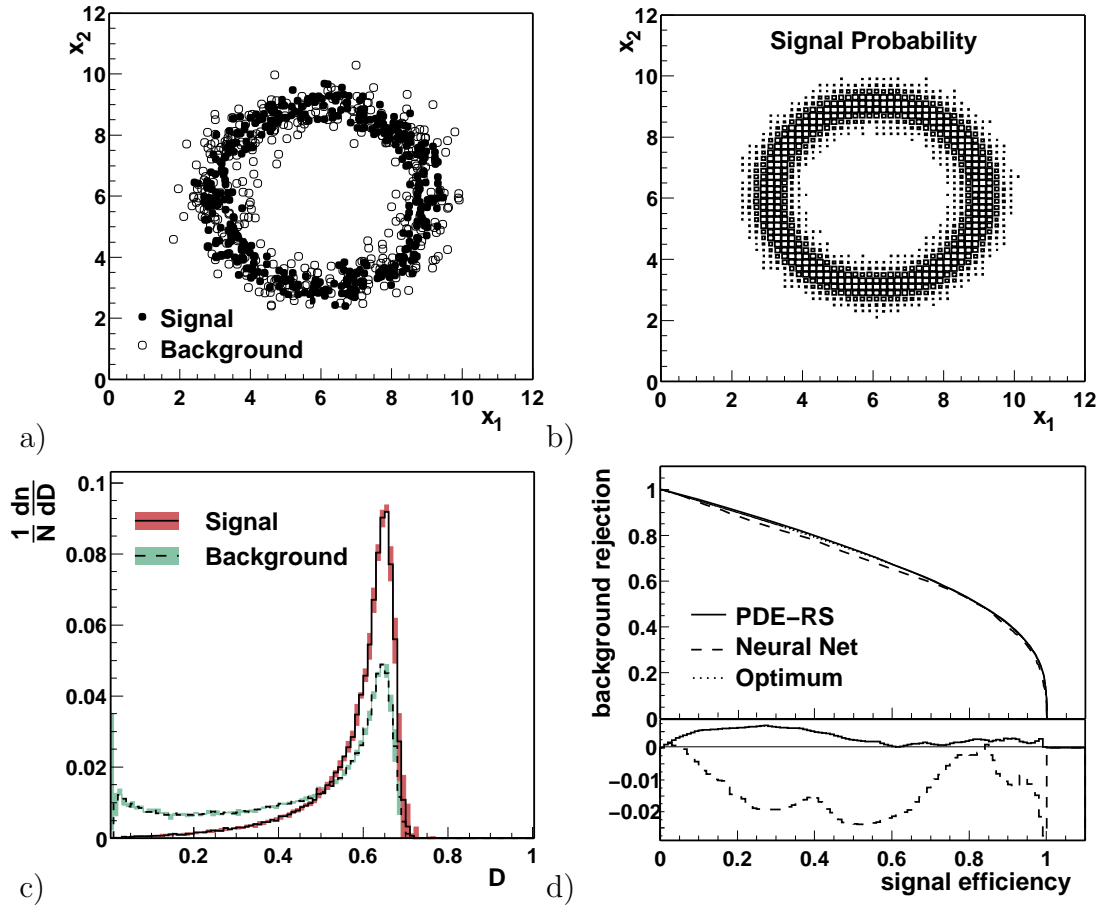}}
    \put(0,6.){ a)}
    \put(7,6.){ b)}
    \put(0,0){ c)}
    \put(7,0){ d)}
  \end{picture}
  \caption{
    \label{fig:ring}
    a) The phase space density of background (open circles) and signal
    (full circles) events generated according to Gaussian smeared
    rings which are highly correlated. b) the signal probability
    density in the $x_1$--$x_2$-plane. c) the resulting shape of the
    discriminant distribution for signal and background events. d)
    background rejection, $1 - \epsilon_b$, versus signal efficiency
    $\epsilon_s$ for the PDE-RS method and a NN, obtained by cutting
    on the discriminant distributions. In the lower part of the figure
    the difference of the PDE-RS method (NN) to the optimum is shown.}
\end{figure}
In a second example we study the performance of the PDE-RS for
strongly correlated input observables. Here, events are generated on a
ring and smeared by a Gaussian. The rings of signal and background
have the same diameter $R=3$ and the width of the Gaussian used to
smear the signal is $\sigma_s=1/2$, while for the background this is
$\sigma_b=1/4$. Such an example was also used in \cite{Towers}, where
it was found that the high correlation of the Cartesian variables
$x_1$ and $x_2$ of the events makes classification very difficult for
NN's\footnote{We find that the performance of the NN is good, if a
  very large number of training cycles is used, i.e. if a very large
  computation time is spent.}. The resulting signal and background
event distributions are shown in figure~\ref{fig:ring}a along with the
resulting signal probability in figure~\ref{fig:ring}b.  The shape of
the discriminant distributions for signal and background is depicted
in figure~\ref{fig:ring}c. A slightly smaller volume $V=0.12\cdot0.12$
compared to the previous example was used, again with 100,000 events
to sample the signal and background distributions, each. In
figure~\ref{fig:ring}d the performance of the PDE-RS method is
compared to a NN which has the same architecture as in the previous
example. This time the integrated area for the PDE-RS method is
slightly larger ($0.708\pm0.031$) than for the Neural Network
($0.691$). 
However, the results of the two methods are compatible.
The NN was again trained with 100,000 events using 10
hidden nodes. Since after a transformation to polar coordinates the
given example reduces to a one-dimensional problem which can be solved
analytically, also the theoretically optimal efficiency curve is
shown. Its integrated area is 0.705. The lower part of
figure~\ref{fig:ring}d shows the difference of the curves for the
PDE-RS (NN) and the optimum. The fact that the PDE-RS method performs
slightly better than the theoretical optimum can be explained by the
statistical uncertainties.

While the performance of the PDE-RS method and the NN is similar, the
time to compute the result is not. For the calculation using the
PDE-RS method 224 seconds were needed on an $800\,{\rm Mhz}$ Linux PC
with a RAM of $256\,{\rm Mbyte}$. The same task took 34.6 hours for
the training of every single NN, and several nets had to be tried
before the right combination of training parameters was found!

\subsection{High Dimensional Example}
\begin{figure}
  \setlength{\unitlength}{1cm}
  \begin{picture}(13,5)
    \put(0,0){\includegraphics[width=\textwidth]{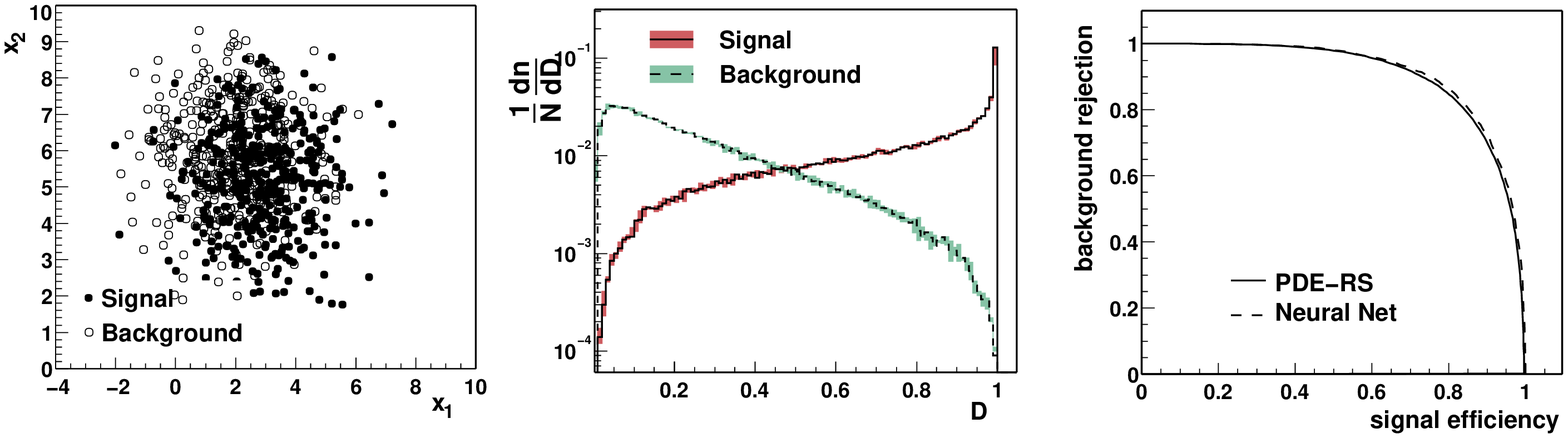}}
    \put(0,0){ a)}
    \put(4.7,0){ b)}
    \put(9.5,0){ c)}
  \end{picture}
  \caption{
    \label{fig:highDim}
    a) Projections of the phase space density for the five dimensional
    example to the observables $x_1$ and $x_2$. The background
    (signal) events are shown as open (full) circles.  b) the
    resulting shape of the discriminant for signal and background
    events. c) background rejection, $1 - \epsilon_b$, versus signal
    efficiency for the PDE-RS method and a NN, obtained by cutting on the
    discriminant distributions.  }
\end{figure}
In this example the behaviour for a large number of observables, i.e.
a problem with high dimensionality is studied. We use a set of five
moderately correlated observables\footnote{ The example is constructed
  as follows: For every signal event a vector $\mathbf
  x'_s = (G(4,1), G(1,1), G(2,1.5), G(2,1), G(1.5,2))$ with components
  sampling normal distributions $G(\left< x\right>, \sigma(x))$ with
  mean $\left< x\right>$ and width $\sigma(x)$ is constructed. This
  vector is then transformed according to
$$ \left( \begin{array}{r}
    x_1 \\
    x_2 \\
    x_3 \\
    x_4 \\
    x_5
\end{array} \right)
= \left( \begin{array}{rrrrr}
    \hphantom{+}1 & -1 & 0& 0& 0\\
    1 & 1 & 0& 0& 0\\
    0 & 0& 1 & 0& 0\\
    0 & 0&\hphantom{+}0& 1 & 1 \\
    0 & 0& 0& -1 & \hphantom{+}1 \\
\end{array} \right) \;
\left( \begin{array}{r}
    x'_1 \\
    x'_2 \\
    x'_3 \\
    x'_4 \\
    x'_5
\end{array} \right)\quad. $$ For background events, the initial vector
is \\
$\mathbf x'_b=(G(4,1), G(2,1), G(3,1.5), G(1,1),
G(0.5,1))$.  } which describe an event. In figure~\ref{fig:highDim}a
the two-dimensional projections on the observables $x_1$ and $x_2$ of
the five-dimensional distribution of signal and background events are
shown. 500,000 events were used to populate the phase space and filled
into the two binary trees. The PDE-RS method can separate signal from
background events using a hypercube with size $l=1.2$. This can be
seen in figure~\ref{fig:highDim}b where the shape of the discriminant
distribution for signal and background events is shown.  The
performance of the NN and PDE-RS methods are compatible. The area under
the background-rejection versus efficiency curve (see
figure~\ref{fig:highDim}c) is $0.906\pm0.008$ for the PDE-RS and 0.910
for the NNs. To get this performance the NN had to be trained with
500,000 events for 1000 training cycles and 10 hidden nodes were used.
If 40 hidden nodes are used and thus 4 times more weights are
available, the area under the background-rejection versus efficiency
curve increases to 0.913 but at the same time the computing time
increased.

\begin{figure}
  \setlength{\unitlength}{1cm}
  \begin{picture}(13,5)
    \put(0,0){\includegraphics[width=6cm,height=5cm]{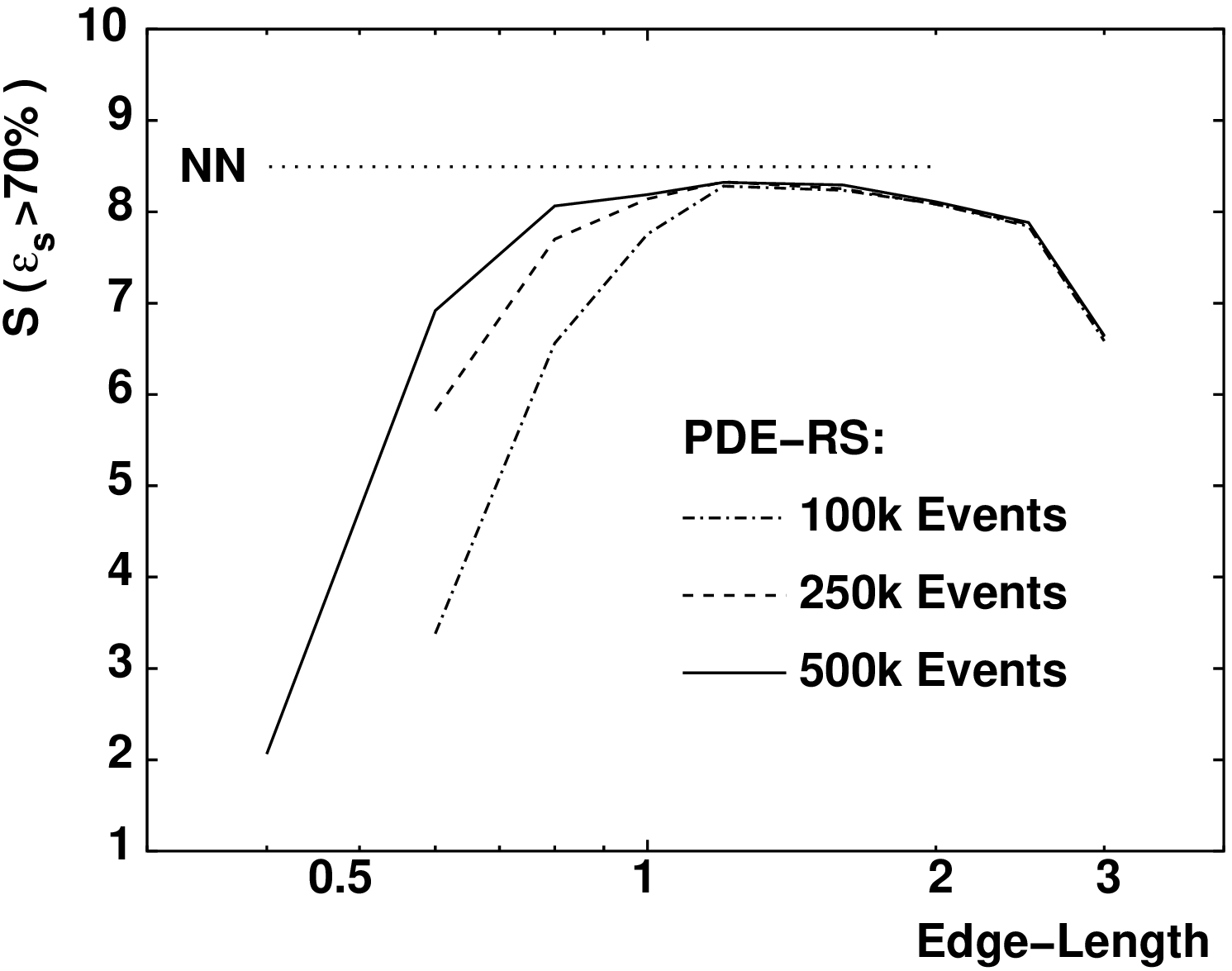}}
    \put(7,0){\includegraphics[width=6cm,height=5cm]{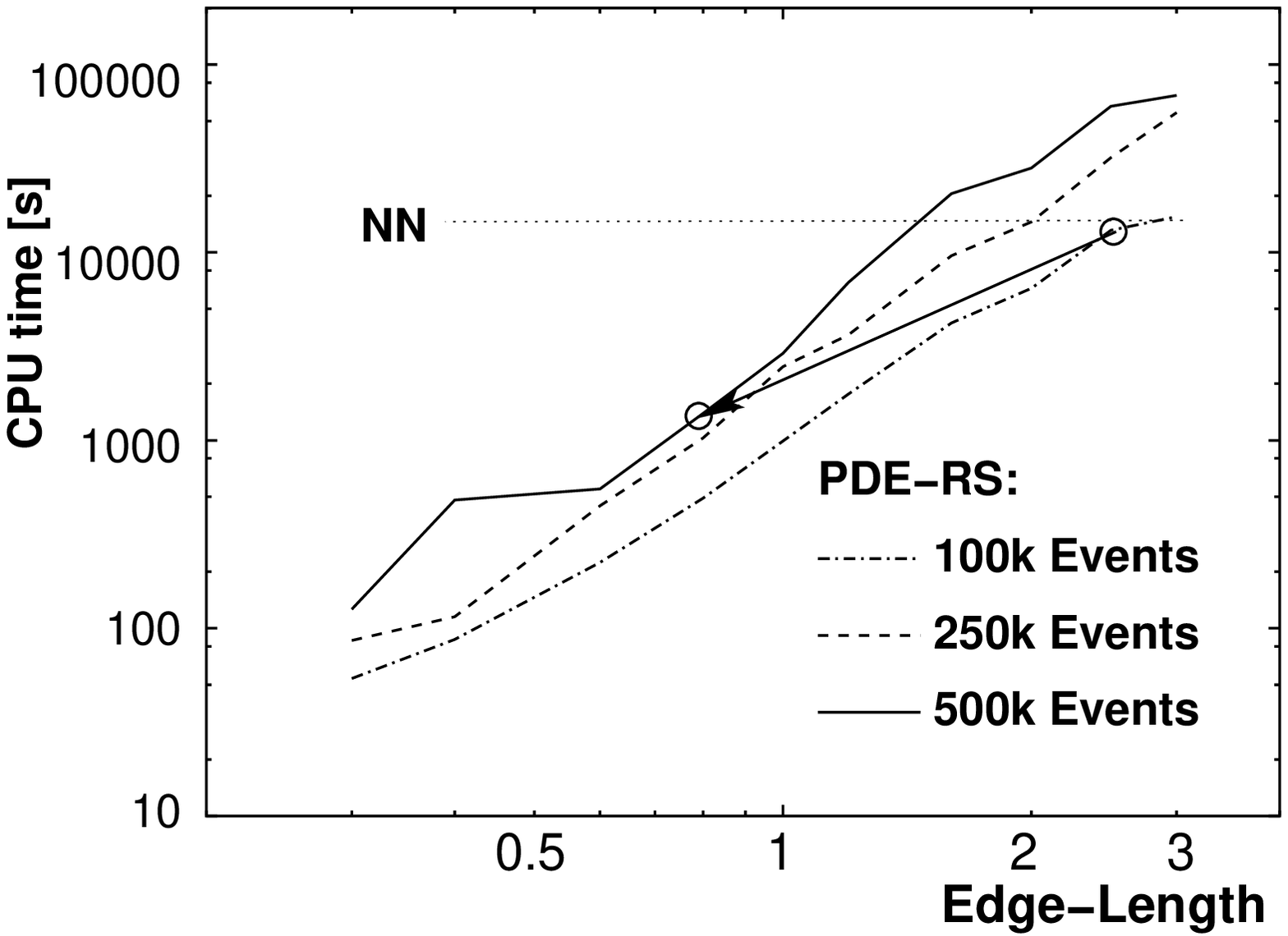}}
    \put(0,0){ a)}
    \put(7,0){ b)}
  \end{picture}
  \caption{
    \label{fig:performance}
    a) The dependence of the separation power, $S = \epsilon_s /
    \epsilon_b$, for fixed signal efficiency on the box size and the
    number of events stored in the binary trees for the
    five-dimensional example, b) computing time needed depending on
    the box size of the PDE-RS method and for the NN. The arrow shows
    how a large numbers of events allow to use smaller box-sizes which
    reduces the computing time needed. }
\end{figure}

Figure~\ref{fig:performance}a shows the separation power
$S:=\epsilon_s/\epsilon_b$ at a signal efficiency of $\epsilon_s=70\%$
as a function of the box size $l$ for the PDE-RS method.  The
separation power has a broad plateau and varies only within 20\%, when
the box size is changed over a large range.  This behaviour
makes the separation power nearly independent of the box-size and will
allow to use the algorithm with a minimum of human
intervention\footnote{The relatively small dependence of the separation power on
  the choice of the box size $l$ has also been verified for the other
  examples discussed here and seems to be a general feature of the
  PDE-RS method.}.

The drop of the separation power towards larger box-sizes is due to
the less accurate mapping of the phase space density to boxes around
the event to be classified. On the other hand, too small boxes will
diminish the number of events in the box and thus will also make the
resolution of the discriminant smaller, because neighbouring events
might end up in different places of the discriminant distribution due
to statistical fluctuations. This can lead to a smearing of
neighbouring events across a larger part of the discriminant.

Figure 5b shows the CPU time needed to compute the PDE-RS results
depending on the box size used and in addition the time needed to
train the NN with 10 nodes. Larger boxes strongly increase the
computing time needed because for all candidate events found by the
binary search within the trees, a time-consuming check needs to be
done whether the event actually falls into the box. The CPU time needed if
more events are used for classification increases logarithmically,
as expected. However, using larger numbers of events for
classification also allows to reduce the box size at the same separation
as indicated by an arrow in figure~5a. 
In the region of good performance of the PDE-RS
method, typically a 10 times smaller computing time than for the NN is
needed.

When comparing the time consumption of the range searching algorithm
to the time needed by a NN, it is interesting to note, that both
algorithms have a very different behaviour. While for the PDE-RS
algorithm the time during the set-up period of filling the binary
trees is more or less the time for reading in the data from a storage
medium, the time needed to train the artificial NN is considerable. On
the other hand, the time needed to classify a single event after the
initialisation phase takes longer with the range searching algorithm,
at least when compared to NNs of moderate size. However, descending
down the binary trees to collect events is a task which naturally can
be done in parallel using multiple threads of program execution.

\section{A Practical Application: Instanton-Induced Processes in DIS at HERA}
\subsection{Instanton-induced Processes in DIS at HERA}
\begin{figure}
  \setlength{\unitlength}{1cm}
  \begin{picture}(10.5,6)
    \put(0,0){\includegraphics[width=5cm]{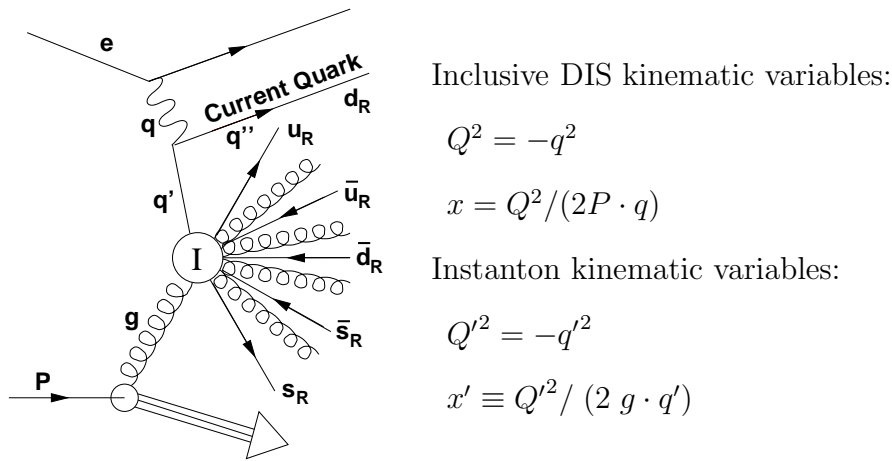}}
    \put(5.5,5){
      \begin{minipage}[t]{8cm}
        Inclusive DIS kinematic variables:
        \begin{description}
        \item{$Q^2=-q^2$}
        \item{$x=Q^2/(2P\cdot q)$}
        \end{description}
        Instanton kinematic variables:
        \begin{description}
        \item{${Q'}^2 = -{q'}^2$}
        \item{$x' \equiv {Q'}^2 / \;(2 \; g \cdot q' ) $}
        \end{description}
      \end{minipage}
    }
  \end{picture}
  \caption{
    \label{fig:instanton}
    Sketch of an instanton-induced process in DIS and the definition
    of the important kinematic variables for inclusive DIS process and
    instanton-induced processes. 
    The four-vector of the
    exchanged photon (incoming proton) is denoted by $q'$ ($P$). The 
    four-vector of the quark (gluon) inducing the hard instanton-process
    is denoted $q'$ (g).
    }
\end{figure}
Instantons \cite{BelavintHooft} are a fundamental non-perturbative
aspect of QCD, inducing hard processes that are absent in perturbation
theory.  The expected cross section in deep-inelastic electron-proton
($ep$) scattering as calculated in ``instanton-perturbation-theory''
\cite{IPert} is sufficiently large to make an experimental discovery
possible \cite{RS1}. However, the background rate is about a factor of
1000 larger --- a challenging task for the classification algorithm.
For a more detailed introduction to instantons-induced processes (\I)
see e.g. \cite{RS4}.

We study the prospect of a search for \I-induced events modelled by
the Monte Carlo simulator QCDINS \cite{QCDINS} which generates
\I-induced events in deep-inelastic $ep$-scattering. In \I-induced DIS
processes (see figure~\ref{fig:instanton}) a quark emerges from a
$q\bar q$-splitting of the exchanged photon and fuses with a gluon
emitted from the proton.  In the \I-induced process $q\bar q$-pairs of
each of the three light quark flavours and on average 2--3 gluons are
produced.  In the hadronic centre-of-mass system (hCMS) they form a
band (of about two units in pseudo-rapidity) of particles with high
transverse energy which are homogeneously distributed in azimuth.
Since in every event a pair of strange quarks is produced, in this
band an increased number of kaons compared to standard DIS events is
expected. Finally, the quark out of the split photon not participating
in the \I-subprocess forms a hard jet. \I-induced events can be
distinguished from standard DIS background events by their
characteristic hadronic final state \cite{RS1,QCDINS,CGRS}. It is therefore
necessary to find hadronic final state observables, which are well
modelled by the background Monte Carlo simulations and which provide
the best possible reduction of the background.

\subsection{Instanton Classification Results}
\begin{figure}[t]
  \includegraphics[width=15cm]{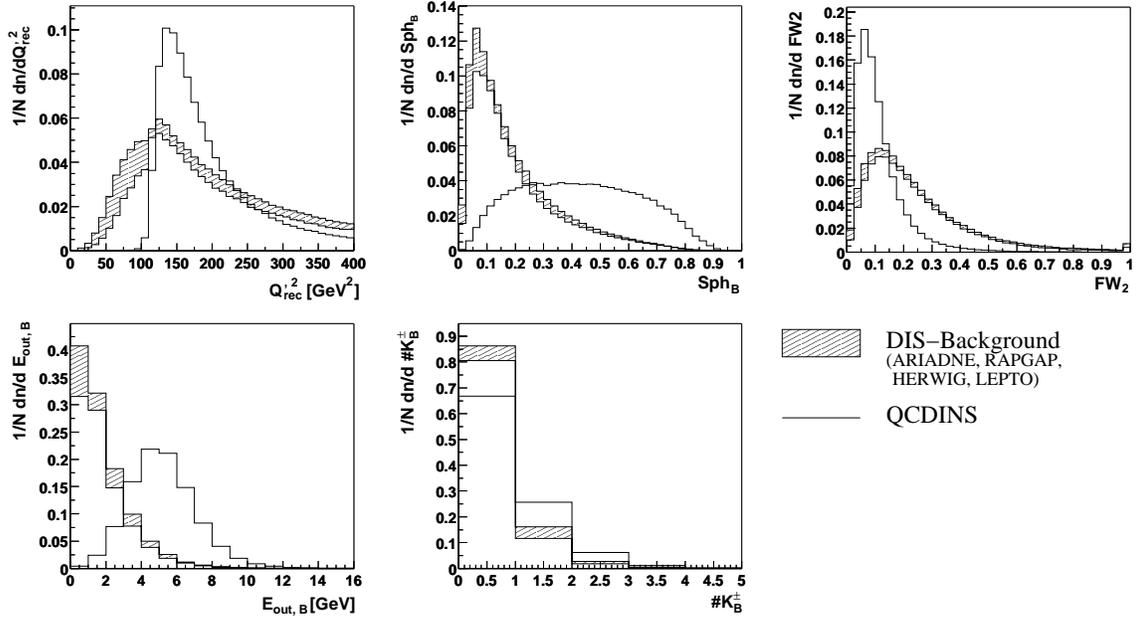}
  \caption{The event observables characterising the hadronic final
  state of $I$-induced processes in DIS at HERA 
 providing good
 instanton separation along with small systematic uncertainties.
 Shown are simulations of $I$-processes (QCDINS) and of standard DIS
 background. The band indicates the uncertainty due to different QCD
 models. The observables are explained in text.  
}
  \label{fig:varsAcat}
\end{figure}
Starting with 35 observables based on the hadronic final state the
best 12 were chosen by calculating the discriminant with all
2-combinations (pairs) of the initial observables and taking those
observables which provide a high separation power
$S=\epsilon_s/\epsilon_b$ demanding an efficiency for instantons of
$\epsilon_s=10\%$. The number of considered observables is further
reduced by calculating all 5-combinations and selecting those with
highest separation power and a small systematic variation of the
background.  The systematic uncertainty was obtained by using four
standard DIS-MC simulators \cite{MCSimulators} which were tuned to
data on representative hadronic final state quantities, in the range
$Q^2>100\GeV^2$ at HERA \cite{MCWorkshop}. The observables forming the
best combination are shown in figure~\ref{fig:varsAcat}.  These are:
the reconstructed virtuality of the quark entering the \I-subprocess
$Q{'}_{\rm rec}^{2}$, the sphericity of the particles in the
\I-Band\footnote{The instanton band is defined to have a width of 2.2
  units of rapidity around the $E_T$-weighted mean rapidity of all
  particles except the jet of the event, taken in the hCMS.} in their
rest system ${\rm Sph}_B$, the second Fox-Wolfram moment \cite{FoxWolfram}
${\rm FW}_2$ of these
particles and the event shape observables $E_{\rm out, B}$ which is
the projection of the particle transverse momentum onto the axis that
makes this quantity maximal \cite{Gibbs} and finally the number of
charged kaons in the \I-Band (see \cite{BKThesis} for a detailed
description of the observables).

\begin{figure}
  \setlength{\unitlength}{1cm}
  \begin{picture}(13,5)
    \put(0,0){\includegraphics[width=6cm,height=5cm]{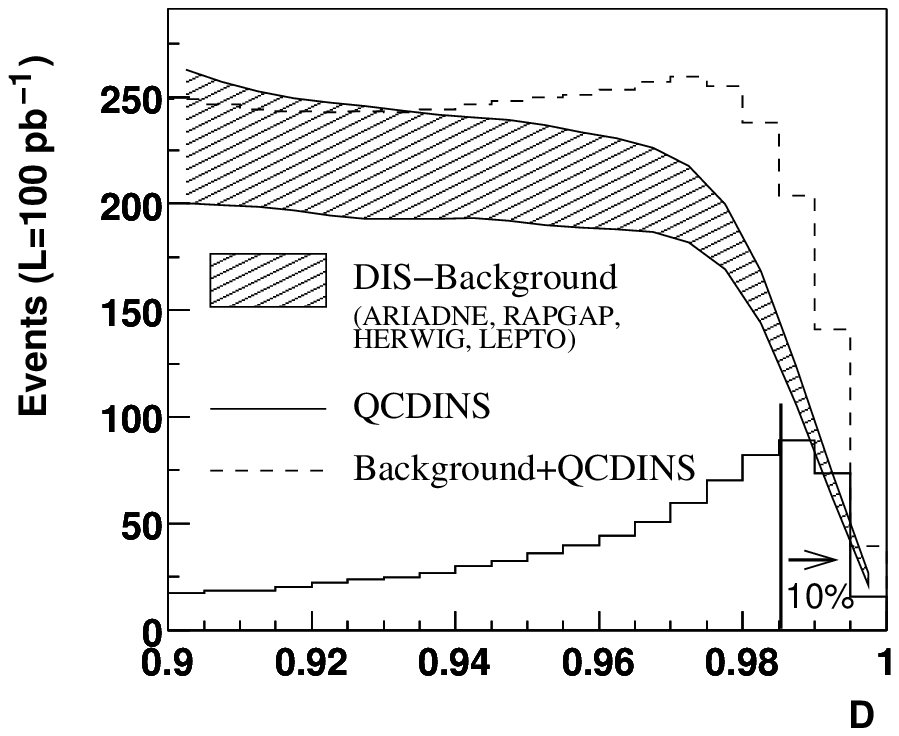}}
    \put(7,0){\includegraphics[width=6cm,height=5cm]{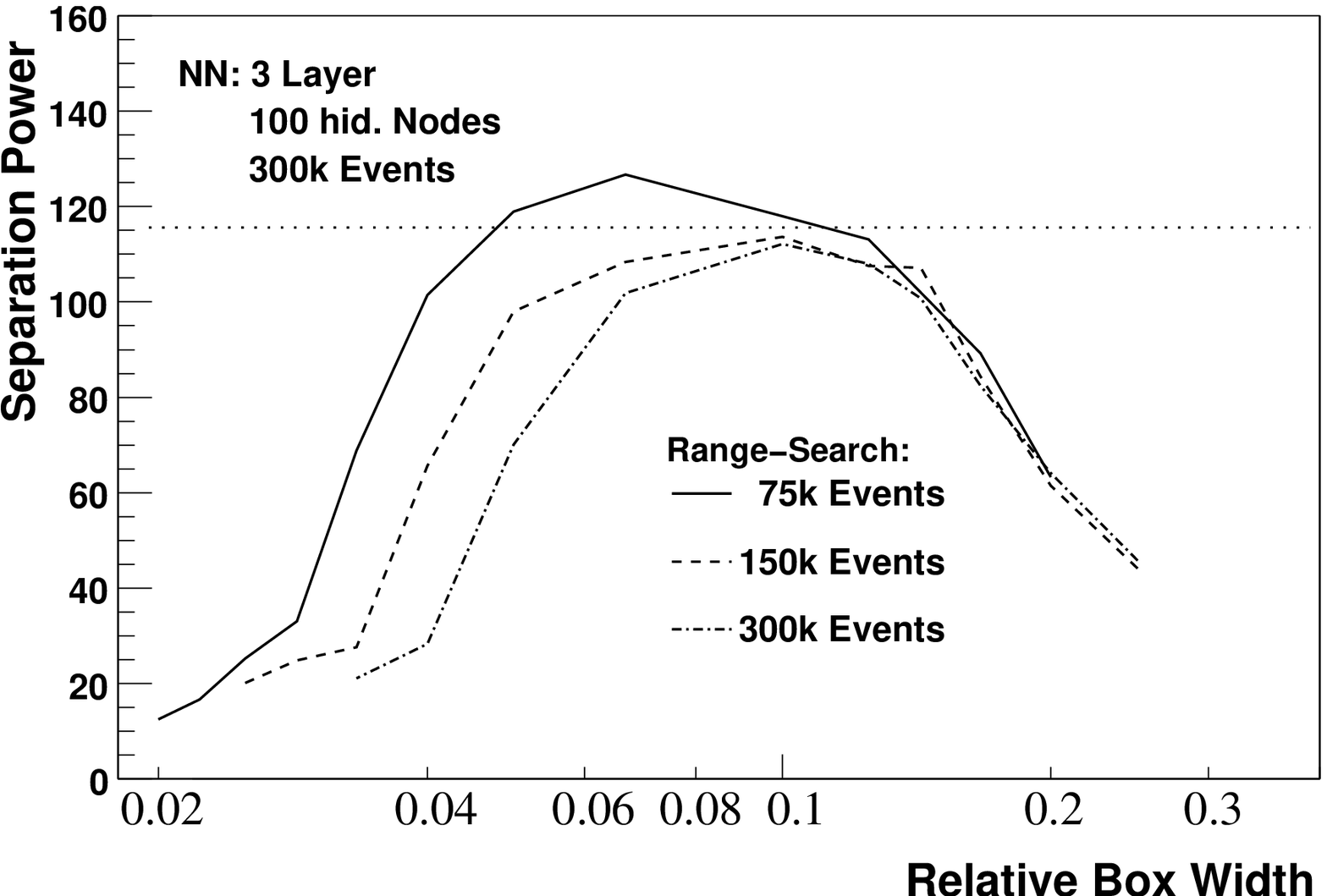}}
    \put(0,0){ a)}
    \put(7,0){ b)}
  \end{picture}
  \caption{a) Number of events expected for an integrated luminosity of $100
    \,{\rm pb}^{-1}$ as a function of the discriminant for instanton-induced
    and standard DIS processes. Only the signal
    region with $D>0.9$ is shown. b) Separation power $S$ at
    $\epsilon_s=10\%$ for different relative box widths and different
    numbers of events in the binary trees. }
  \label{fig:resultsAcat}
\end{figure}

The separation power for $\epsilon_s=10\%$ is $S=126$. In
figure~\ref{fig:resultsAcat}a the expected number of \I-induced events
in DIS at HERA and the number of background events is shown as a
function of the discriminant $D$, for a data sample of an
integrated luminosity of $100\,{\rm pb}^{-1}$. Only the signal region
defined by $D > 0.9$ is shown.  The luminosity is comparable to the
one already collected by each of the HERA experiments H1 and ZEUS. An
event sample can be isolated where half of the events are instantons
while the \I-efficiency is still 10\%. The decreasing background model
uncertainty in the signal region reflects the choice of observables
with a minimum background uncertainty, which was possible due to the
speed and flexibility of the PDE-RS method.

To reduce the number of parameters for the box size, the ratios of the
box edge lengths were fixed by defining a box which contains most of
the events and letting $V$ be a scaled version of this large box. The
projections onto these box edges are shown in
figure~\ref{fig:varsAcat}. The variation of the result depending on
the size of $V$ is shown in figure~\ref{fig:resultsAcat}b. The
behaviour is similar to the one in the toy-model: The separation
increases for smaller boxes with the number of events that populate
the search trees, while for larger boxes this difference vanishes. The
width of the plateau increases with the number of events in the
classification tree. If the data sample is large enough, the width of
the plateau spans one order of magnitude.  This allows to use only an
approximate size parameter which reduces the need for fine tuning, if
a large enough Monte Carlo data sample is available.

In addition a comparison with a single hidden layer feed forward NN
was done. Several network architectures were tried. The network
performing best had 3 layers with 100 hidden nodes.  It reached a
separation of $S=116$ at an \I-efficiency of 10\%, being slightly
worse than the PDE-RS method. To get this good performance the high
numbers of nodes was mandatory.  Training the net was rather time
consuming\footnote{ $4\,{\rm h}$ compared to $20\,{\rm min}$ for the
  PDE-RS method on an $800\,{\rm Mhz}$ Linux PC with a RAM of
  $256\,{\rm Mbyte}$.}  and a lot of human intervention was needed to
adjust the training parameters.

\section{Conclusions}
For the examples covering different basic problems appearing in High
Energy Physics data analyses the presented new classification
algorithm based on range searching (PDE-RS) has a discrimination power
which is comparable to the one provided by Neural Networks. However,
the PDE-RS method needs less computation time, is more transparent and
is rather insensitive to the choice of the free parameters that need
to be set manually.  Moreover the classification error can be easily
evaluated.  For complex cases the speed of the algorithm allows to
carefully choose the best combination of observables which give the
best separation and for which the observables and their correlations
are well described by the simulations. The PDE-RS method is therefore
a powerful tool to find a small number of signal events in the large
data samples expected at future colliders. It is particularly suited
for hadron colliders where the background is large and the correct
background description is difficult.

\section{Acknowledgements}
We would like to thank Prof.~V.~Blobel for making us aware of the
range search algorithms and for many discussions helpful to develop a
method suitable for High Energy Physics applications. A.~Ringwald and
F.~Schrempp we would like to thank for many years of fruitful
collaboration on investigating search strategies for \I-induced
processes in DIS. Many thanks also for initiating
first ideas to apply multi-variate discrimination techniques 
for \I-searches. 
Finally we would like to express our
gratitude to P.~C.~Bhat, V.~Blobel, L.~Holmstr\"om, 
C.~Kiesling, D. Lelas and J.~Zimmermann for
comments on the manuscript.


\end{document}